\documentclass[aps,prb,twocolumn,showpacs,amsmath,amssymb]{revtex4}
\usepackage{epsfig}
\usepackage{dcolumn}
\usepackage{bm}
\usepackage{amsmath}
\usepackage{amsfonts}
\usepackage{amssymb}
\usepackage{graphicx}%
\newcommand{\beq}{\begin{eqnarray}}
\newcommand{\eeq}{\end{eqnarray}}

\begin{document}

\title{Stationary transport in mesoscopic hybrid structures with
 contacts to superconducting and normal wires.
A Green's function approach for multiterminal setups.}
\author{Liliana Arrachea}
\affiliation{Departamento de F\'{\i}sica, Facultad de Ciencias Exactas y Naturales, 
Universidad de Buenos Aires, Pabell\'on I, Ciudad Universitaria, 1428 
Buenos Aires, Argentina.\\
Departamento de F\'{\i}sica de la Materia Condensada and BIFI, Universidad
de Zaragoza,  Pedro Cerbuna 12, 50009 Zaragoza, Spain.}
\begin{abstract}
We generalize the representation of the real time Green's functions
introduced by Langreth and Nordlander 
[Phys. Rev. B {\bf 43} 2541 (1991)] and Meir and Wingreen
[Phys. Rev. Lett. {\bf 68} 2512 (1992)]  in stationary quantum transport
in order to  study problems with hybrid structures containing 
normal (N) and superconducting (S) pieces. We illustrate the treatment
in a S-N junction under a stationary bias and investigate in detail the
behavior of the equilibrium currents in a normal ring threaded by a 
magnetic flux with attached superconducting wires at equilibrium.
We analyze the flux sensitivity of the Andreev states and we show that
their response is equivalent to the one corresponding to the Cooper
pairs with momentum $q=0$ in an isolated 
superconducting ring.
\end{abstract}
\maketitle

\section{Introduction.}
The superconductivity and its implications is among the most interesting phenomena 
in the realm of condensed matter physics. While the microscopic mechanism leading to the pairing instability
in the high-$T_c$ materials remains not yet fully understood, the general framework provided by the BCS theory \cite{BCS} 
consistently 
accounts for superconductivity in normal metals. Remarkably, this seems to be even true in the context
of low dimensional systems of mesoscopic scale. \cite{rev,exp}

 BCS theory provided the basis of the seminal paper by Blonder, Tinkham and Klapwijk  (BTK) \cite{btk}.
In that work, 
the stationary transport properties of a superconductor-normal  metal (S-N) junction and the subtle mechanism of the
Andreev 
reflection leading to the effective Cooper pair tunneling through the junction was first analyzed. A similar
description was followed in  the study of  S-N-S structures, \cite{ringbu,ringcay,grambu,aflec,been} and
 later formulated in terms of multichannel scattering matrix theory in Ref \onlinecite{beenan}.
  BCS theory has been also
the basis for the study of stationary transport in unbiased S-N-S junctions due to the 
Josephson effect 
\cite{BCS,mahan,alf,sols,vec} as well as the AC Josephson effect under bias \cite{BCS,mahan,cue,claro,melin}.

The non-equilibrium Green's function formalism \cite{negf} is a powerful technique
to study quantum transport in coherent regimes. In the context of microscopic models for mesoscopic structures
it was first introduced by Caroli {\em et al}, \cite{caroli} and later
elaborated by other authors. \cite{file,langreth,past,meiwi1,meiwi2,lilip} That approach  was also represented in the Nambu 
formalism to treat 
 S-N  and S-N-S junctions. \cite{alf,cue,claro,melin}
 The formal equivalence between  non-equilibrium 
Green's function and the scattering matrix formalism to the quantum transport 
has been analyzed for the case of
normal systems without many-body interactions under stationary \cite{file} and 
time-periodic driving. \cite{lilip} 

The representation of the non-equilibrium Green's functions introduced
by Langreth and Nordlander \cite{langreth} is particularly useful to derive compact 
equations for the currents along the different pieces of a mesoscopic structure. \cite{meiwi1,meiwi2}
 In the present work, we employ that representation  
in the case of hybrid multiterminal structures 
containing superconducting elements that are modeled by BCS Hamiltonians. 

Instead of working in  Nambu's space,
we derive a coupled set of Dyson's equations for the normal 
$\hat{G}^{R,<}_{\sigma}(\omega)$ and  Gorkov's 
$\hat{F}^{R,<}_{\sigma}(\omega)$ retarded ($R$) and lesser ($<$) Green's functions. 
As in Refs.  \cite{past,meiwi1,meiwi2}, we ``integrate- out'' the degrees
of freedom of the external wires (reservoirs) and, by introducing auxiliary hole propagators 
$\hat{\overline{g}}^{R, <}(\omega)$, we reduce the problem to solving the Dyson's equation
for the usual normal Green's function with an effective self-energy. As in  
Refs.  \cite{past,meiwi1,meiwi2}, the latter
describes the scattering events due to the escape to the leads, but in the present case,
it contains a component related to the multiscattering processes involved in the Andreev 
reflection.
The final expressions for the currents have a compact structure that
formally resemble those
of Ref. \onlinecite{meiwi1} for normal systems. 

Sections II and III are devoted to explain the 
theoretical treatment.
We derive  expressions for the currents and we show that the transmission function of a biased 
system
contains a normal plus an Andreev contribution.
In Section IV we illustrate the approach in the simple well known case of a two terminal setup 
with
a linear system in contact to one normal and 
one superconducting wires under bias and we show its equivalence with BTK description. 
 In Section V we employ the formalism to the study of  the behavior of the
equilibrium currents of a normal metallic ring threaded
by a static magnetic field 
 with several attached normal and/or superconducting wires. We address several interesting physical
questions like  the minimal conditions for the development of Andreev states within the superconducting gap,
the flux sensitivity of these states and the possibility of anomalous flux quantization induced as a consequence
of the proximity effect.
Section VI is devoted to summary and discussion. Some technical details are 
presented in the appendices.

\section{Theoretical treatment.}
\begin{figure}
\includegraphics[width=0.9\columnwidth,clip]{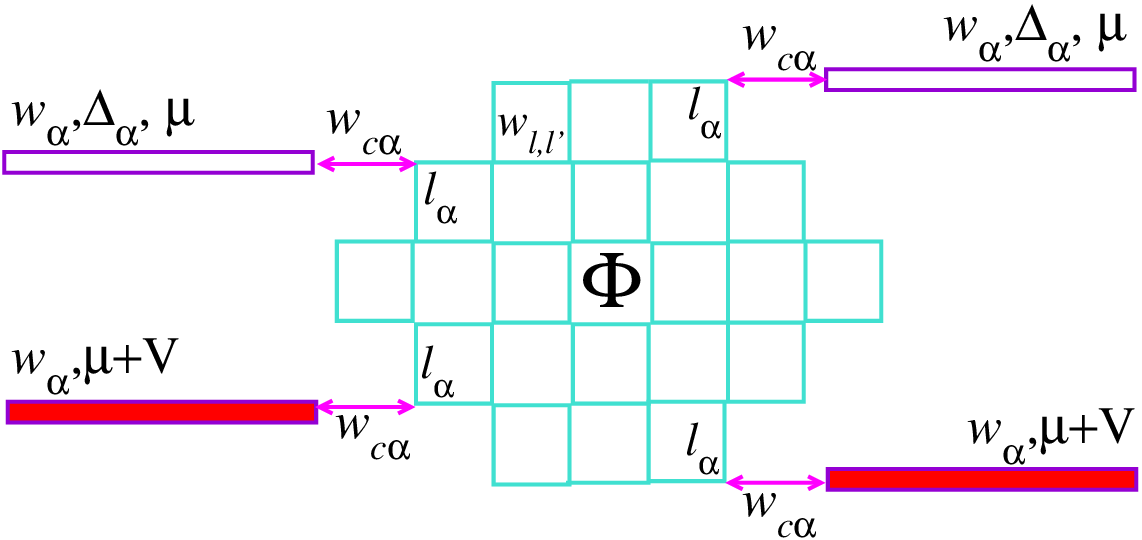}
\caption{\label{fig1} (Color online) Sketch of the setup. The central grid represents
the central finite system. The area enclosed by this system is threaded by a static magnetic flux 
$\Phi$. The N- and  S wires  are,
respectively, indicated with open and filled lines.
  The arrows represent
the contacts between the different systems.
 In each case, the parameters
of the ensuing Hamiltonians are indicated.}
\end{figure}

\subsection{Model}
We introduce microscopic models for the different pieces of the setup,
which consists in a finite normal system of non-interacting electrons 
in contact to 
$M$ infinite superconducting (S) or normal (N) metallic wires (see Fig. 1). The full system  is
described by the following Hamiltonian:
\begin{equation}
H=H_{cen}+ \sum_{\alpha=1}^{M} (H_{\alpha} + H_{c \alpha} ),
\end{equation}
where $H_{\alpha}$ 
denote the Hamiltonians
of the  wires, while $ H_{c \alpha}$ the corresponding
contacts establishing the connections between these systems and the  central one.
Although long-range superconducting order does not take place in strictly one-dimension (1D),
for simplicity, we consider 1D tight-binding BCS Hamiltonians with local
s-wave pairing for the wires. This is a rather standard assumption (see
Refs. \onlinecite{btk,ringbu,ringcay,grambu,aflec,alf,vec,cue,claro,melin}) and
 the general treatment can be easily extended to
multichannel wires and more general symmetries of the superconducting gap.
Concretely:
\begin{eqnarray}
H_{\alpha}&=&-w_{\alpha} \sum_{j_{\alpha}=1,\sigma }^{L_{\alpha}} 
(c^{\dagger}_{j_{\alpha}, \sigma}c_{j_{\alpha}+1, \sigma} + H. c. ) - 
 \\
& &
\mu_{\alpha} \sum_{j_{\alpha}=1,\sigma }^{L_{\alpha}}
c^{\dagger}_{j_{\alpha}, \sigma}c_{j_{\alpha}, \sigma} +
\sum_{j_{\alpha}=1}^{N_{\alpha}} 
(\Delta_{\alpha}  c^{\dagger}_{ j_{\alpha},  \uparrow} 
c^{\dagger}_{j_{\alpha}, \downarrow} + H.c. ),\nonumber
\end{eqnarray}
with $\sigma=\uparrow,\downarrow$, and
being $\Delta_{\alpha}=0$ for the N-wires.
 The size of the wires approaches the thermodynamic
limit  ($L_{\alpha} \rightarrow \infty$), i.e., the wires act as
macroscopic reservoirs, with well defined chemical potential and temperature.
We model the central system by a  
tight-binding Hamiltonian in a finite lattice of
$L$ sites with nearest-neighbor hopping. We consider the possibility of a static magnetic flux $\Phi$ threading this
system, which introduces a dependence on $\Phi$ in the hopping matrix elements:
\begin{eqnarray}
H_{cen}&= &- \sum_{ \langle l l'\rangle ,\sigma} [w_{l,l'}(\Phi)
c^{\dagger}_{l, \sigma} c_{l', \sigma} + H.c.] \nonumber \\
& & + \sum_{ l=1 ,\sigma}^L \varepsilon^0_l c^{\dagger}_{l, \sigma} c_{l, \sigma},
\end{eqnarray}
where $\langle l l'\rangle $ denotes nearest-neighbor sites.
The Hamiltonians for the contacts read:
\begin{equation}
H_{c \alpha} = -w_{c\alpha}  \sum_{\sigma}
(c^{\dagger}_{ j_{c \alpha },  \sigma} c_{l_{c  \alpha}, \sigma} + H.c.),
\end{equation}
which describe hopping processes between the sites $j_{c \alpha}$ of the 
wires and the sites $l_{c \alpha}$
of the central system at which the wires are attached.

\subsection{Currents.}
The electronic current, in units of $e/\hbar$, flowing through a given bond $\langle l, l' \rangle$
of the central system  is:
\begin{eqnarray}\label{curing}
J_{l,l'} 
& = & 
-2  \sum_{\sigma}
\int_{-\infty}^{\infty} \frac{d \omega}{2 \pi}
\mbox{Re}[w_{l',l}(\Phi) G^<_{l,l',\sigma}( \omega)],
\end{eqnarray}
while the current flowing through a given contact is
\begin{eqnarray}\label{curwi}
J_{\alpha} 
& = & 
-2 \sum_{\sigma}
\int_{-\infty}^{\infty} \frac{d \omega}{2 \pi}
\mbox{Re}[w_{c \alpha} G^<_{j_{c \alpha },l_{c \alpha},\sigma}( \omega)],
\end{eqnarray}
being 
\begin{eqnarray}
G^<_{l,l',\sigma}(t,t') & = &i \langle c^{\dagger}_{l \sigma}(t) 
c_{l' \sigma}(t') \rangle,
\end{eqnarray}
and $G^<_{l,l',\sigma}(\omega)$ the corresponding Fourier transform in $t-t'$. 

\subsection{Evaluation of the Green's functions.}
In previous literature, the evaluation of the Green's functions for
hybrid structures described in terms of tight-binding and BCS 
Hamiltonians has been carried out in the framework of the Nambu formalism
\cite{alf,cue,claro,melin}. We briefly present bellow an alternative and
equivalent representation, which will allow us to analyze from a different perspective
the physical 
processes involved in the phenomena of Andreev reflection and the development of
Andreev states within the superconducting gap.

We define retarded normal and Gor'kov Green's functions:
\begin{eqnarray}
& & G^R_{j,j',\sigma}(t,t')  =  -i \Theta(t-t') 
\langle \{ c_{j, \sigma}(t), 
c^{\dagger}_{j', \sigma}(t') \} \rangle, \nonumber \\
& & F^R_{ j,j',\sigma}(t,t')  = -i  \Theta(t-t')
 \langle \{ c^{\dagger}_{j, \sigma}(t), 
c^{\dagger}_{j', \overline{\sigma}}(t') \}  \rangle,
\nonumber
\end{eqnarray}
where $\{.,. \}$ denotes the anticommutator of the corresponding operators and 
 $\overline{\uparrow}=\downarrow$, $\overline{\downarrow}=\uparrow$.

It can be verified that the equations of motion for these functions
are coupled and read:
\begin{eqnarray}
& & \omega G^R_{j,j',\sigma}(\omega) - \sum_{j''}
\varepsilon_{j,j''}  G^R_{j'',j',\sigma}(\omega)
- \Delta_j F^R_{j,j',\sigma}(\omega) = \delta_{j,j'}\nonumber \\
& & \omega F^R_{j,j',\sigma}(\omega) + \sum_{j''}
\varepsilon_{j,j''}  F^R_{j'',j',\sigma}(\omega)
- \Delta^*_j G^R_{j,j',\sigma}(\omega) = 0.
\nonumber
\end{eqnarray}
The spacial indexes extend over the coordinates of the whole system.
For coordinates on the wires 
$\varepsilon_{j,j'} = \sum_{\alpha}  \delta_{j,j_{\alpha}} 
(\delta_{j,j'} \mu_{\alpha} -  \delta_{j\pm 1,j'} w_{\alpha})$, 
$\Delta_j =  \sum_{\alpha} \Delta_{\alpha} \delta_{j,j_{\alpha}}$.
For coordinates on the central system: $\varepsilon_{j,j'}= - \sum_{\langle l,l' \rangle}
\delta_{j,l} \delta_{j',l'} w_{l,l'}(\Phi)$,
for $\langle l,l' \rangle$, 
being nearest neighbors within the $L$-site lattice, 
$\varepsilon_{j,j'}= \sum_{l=1}^L \varepsilon^0 \delta_{l,j} \delta_{j,j'}$
and $\Delta_j = 0$. For coordinates on the contacts:
 $\varepsilon_{j,j'}=-
w_{c \alpha} (\delta_{j,l_{c \alpha}} \delta_{j', j_{c \alpha} }+
\delta_{j,j_{c \alpha }} \delta_{j',l_{c \alpha}} )$ and $\Delta_l =0$.

As usual, it is convenient to eliminate the degrees of freedom
of the wires. Such a procedure defines self-energies for the Green's
functions with coordinates belonging to what we have defined as the central
system. \cite{meiwi1,meiwi2} We summarize it in Appendix \ref{appintout} for the present problem.
The result is that the retarded Green's functions
with coordinates on the central system can be expressed as elements
of $L \times L$ matrices and the ensuing Dyson's equations read:
\begin{eqnarray} \label{retc}
& & [\hat{g}^R(\omega)]^{-1}\hat{G}^R_{\sigma}(\omega)
+ \hat{\Sigma}^{gf, R} (\omega) \hat{F}^R_{\sigma}(\omega)  = \hat{1}, \nonumber \\
& &  [\hat{\overline{g}}^R(\omega)]^{-1} \hat{F}^R_{\sigma}(\omega)
+ \hat{\Sigma}^{fg, R} (\omega) \hat{G}^R_{\sigma}(\omega)  = \hat{0},
\end{eqnarray}
where $\Sigma_{l,l'}^{\nu \nu', R}(\omega)=\delta_{l,l'} 
\sum_{\alpha} \delta_{l, l_{c \alpha}}
\Sigma_{\alpha}^{\nu \nu',R}(\omega)$, with $\nu,\nu'=g,f$. The explicit evaluation of these 
functions
is summarized in Appendix B. The have introduced the retarded Green's functions
$\hat{g}^R(\omega)$ and  $\hat{\overline{g}}^R(\omega)$, whose corresponding
inverses are:
\begin{eqnarray} \label{overg}
& & [\hat{g}^R(\omega)]^{-1}  =  
\omega \hat{1} - \hat{\varepsilon}(\Phi)- \hat{\Sigma}^{gg, R}(\omega), \nonumber \\
& & [ \hat{\overline{g}}^R(\omega)]^{-1}  = 
\omega \hat{1} + \hat{\varepsilon}(-\Phi)- \hat{\Sigma}^{ff, R}(\omega),
\end{eqnarray}
where $\hat{\varepsilon}(\Phi)$ contains the matrix elements of the Hamiltonian $H_{cen}$. 
In the case that all the wires are normal 
($\Delta_{\alpha}=0, \forall \alpha$), the function $\hat{g}^R(\omega)$
is the exact retarded normal Green's function of the coupled central system, while 
$\Sigma^{ff, R}(\omega)= -[\Sigma^{gg, R}(-\omega)]^*$, 
thus  $\hat{\overline{g}}^R(\omega)=[\hat{g}^R(-\omega)]^*$,
which indicates that  $\hat{\overline{g}}^R(\omega)$ is a  propagator related to the dynamics of
the holes.

The second equation (\ref{retc}) can be casted:
\begin{equation}\label{f}
\hat{F}_{\sigma}^R(\omega)=-\hat{\overline{g}}^R(\omega)\hat{\Sigma}^{fg, R}(\omega) 
\hat{G}_{\sigma}^R(\omega).
\end{equation}
Substituting (\ref{f}) in the first equation (\ref{retc}) the formal solution for the 
normal Green's
is obtained:
\begin{equation}\label{gret}
[\hat{G}_{\sigma}^R(\omega)]^{-1}=\omega \hat{1}- \hat{\varepsilon} - \hat{\Sigma}^R_{eff}(\omega) ,
\end{equation}
where we have defined an {\em effective normal self-energy}:
\begin{equation}\label{efself}
\hat{\Sigma}^R_{eff}(\omega)=\hat{\Sigma}^{gg, R}(\omega)+
\hat{\Sigma}^{gf, R}(\omega) \hat{\overline{g}}^R(\omega) \hat{\Sigma}^{fg, R}(\omega).
\end{equation}

The lesser counterpart of (\ref{gret}) is, thus, written as:
\begin{equation}\label{gless}
\hat{G}^<_{\sigma}(\omega)= \hat{G}^R_{\sigma}(\omega) \hat{\Sigma}_{eff}^< (\omega) 
\hat{G}^A_{\sigma}(\omega),
\end{equation}
being the advanced Green's function $ \hat{G}^A_{\sigma}(\omega)
=[ \hat{G}^R_{\sigma}(\omega)]^{\dagger}$
Using Langreth rules \cite{langreth}: $(BC)^<= B^R C^<+ B^< C^A$ in the definition of
(\ref{efself}), it can be shown that
\begin{eqnarray}\label{sigefless}
\hat{\Sigma}_{eff}^< (\omega)& & =\hat{\Sigma}^{gg,<} (\omega) +\nonumber \\
& & 
\hat{\Sigma}^{gf, <} (\omega) \hat{\overline{g}}^A (\omega)  
\hat{\Sigma}^{fg, A} (\omega) + \hat{\Sigma}^{gf, R} (\omega) \nonumber \\
& & \times [ \hat{\overline{g}}^<(\omega)  
\hat{\Sigma}^{fg, A} (\omega)+ \hat{\overline{g}}^R(\omega)  
\hat{\Sigma}^{fg, <} (\omega)].
\end{eqnarray}
Using the lesser counterpart of (\ref{overg}):
\begin{equation}
\hat{\overline{g}}^<(\omega)=
\hat{\overline{g}}^R(\omega) \hat{\Sigma}^{ff, <} (\omega) \hat{\overline{g}}^A(\omega),
\end{equation}
the lesser effective self-energy $\hat{\Sigma}_{eff}^< (\omega)$ can be fully expressed
in terms of the bare ones, $\Sigma^{\nu \nu', <}_{\alpha} (\omega)=
i f_{\alpha}(\omega) \hat{\Gamma}^{\nu,\nu'}_{\alpha}(\omega)$, 
with $\nu,\nu'= g,f$, which depend on the temperature 
$T_{\alpha}$ of the reservoirs through the
Fermi function $f_{\alpha}(\omega)$: 
\begin{eqnarray}\label{sigefles}
\Sigma^<_{eff, \alpha, \beta}(\omega) &=&
\delta_{\alpha,\beta} \Sigma^{gg, <}_{\alpha} (\omega)
+\nonumber \\
& & \Lambda^R_{\alpha,\beta}(\omega) 
 \Sigma^{fg, <}_{\beta} (\omega) + \Sigma^{gf, <}_{\alpha} (\omega)
\Lambda^A_{\alpha,\beta}(\omega) 
\nonumber \\
& &+ \sum_{\alpha'}\Lambda^R_{\alpha,\alpha'}(\omega) \Sigma^{ff, <}_{\alpha'} (\omega)
\Lambda^A_{\alpha',\beta}(\omega),
\end{eqnarray}
with $\Lambda^R_{\alpha,\beta}(\omega) =
\Sigma^{gf,R}_{\alpha}(\omega) \overline{g}^R_{l_{c\alpha},l_{c\beta}}(\omega)$ and
$\Lambda^A_{\beta, \alpha}(\omega)= [\Lambda^R_{\alpha,\beta}(\omega)]^*$.
Alternatively, the above expressions can be also directly obtained after some algebra
from the lesser counterpart of (\ref{retc}), as indicated in Appendix C.

At equilibrium, it is satisfied:
\begin{equation}\label{sieq}
\Sigma^<_{eff,\alpha,\beta}=i f(\omega)\Gamma_{eff,\alpha,\beta}(\omega),
\end{equation}
being
\begin{eqnarray}\label{gamaef}
\Gamma_{eff,\alpha,\beta}(\omega) & = &i [\Sigma^R_{eff,\alpha,\beta}(\omega)-
\Sigma^A_{eff,\alpha,\beta}(\omega)] 
 =\nonumber \\
& &
\delta_{\alpha,\beta}  \Gamma^{gg }_{\alpha} (\omega)
 + \Lambda^R_{\alpha,\beta}(\omega) 
 \Gamma^{fg }_{\beta} (\omega) + \nonumber \\
& & \Gamma^{gf}_{\alpha} (\omega)
\Lambda^A_{\alpha,\beta}(\omega) + 
\nonumber \\
& &  \sum_{\alpha'=1}^M\Lambda^R_{\alpha,\alpha'}(\omega) \Gamma^{ff}_{\alpha'} (\omega)
\Lambda^A_{\alpha',\beta}(\omega),
\end{eqnarray}
which implies:
\begin{equation}\label{geq}
G^<_{l,l',\sigma}(\omega)= f(\omega)[G^A_{l,l',\sigma}(\omega)-G^R_{l,l',\sigma}(\omega)].
\end{equation}

Before closing this section, let us emphasize the formal equivalence
between Eqs. (\ref{gret}) and (\ref{gless}) and the 
 representation of Ref. \onlinecite{langreth,meiwi1,meiwi2}. 
In the present case, the effective self-energies (\ref{efself}) and
(\ref{sigefles}), however, have a more complicated structure
when the leads are superconducting. In particular, they contain the normal terms 
$\hat{\Sigma}^{gg,R,<} (\omega)$ that represent the normal ``escape to the leads'' of single
electrons, as well as terms involving {\em multiple scattering processes}, mediated by the
hole propagators $\hat{\overline{g}}^{R,<}(\omega)$. The latter
act not only locally, but
also extend along the different positions of the sample that are in contact to 
superconducting wires.

\section{Stationary currents and transmission functions.}
Being able to evaluate the lesser Green's functions, 
we are now in the position to evaluate the currents (\ref{curing}) and
(\ref{curwi}).
We recall that a biased setup with several superconducting wires defines, in general,
a time-dependent problem.\cite{cue,scal} In this work we are interested in
the stationary transport. Thus, in what
follows we shall derive expressions for the currents in two
situations: (i) A biased setup with a voltage difference between the S-
and the N- wires, being all the S wires at the same chemical potential.
In this case, currents flow through the contacts as well as along 
the central system. 
  (ii) The second situation corresponds to all the  
wires at the same chemical potential, in which case, there are no currents
flowing through the contacts and there exists only  the possibility of equilibrium currents
along the central structure
when it is threaded by a finite magnetic flux. We present below general exact
expressions for the currents and we shall address separately the two different 
cases
in the next two sections.

Using Dyson's equation for the lesser Green's function, the expressions
(\ref{curing}) and (\ref{curwi}) cast:
\begin{eqnarray}
J_{l,l'} &= & - 2 \sum_{\sigma, \alpha, \beta =1}^M \int_{-\infty}^{+\infty}
\frac{ d\omega}{2 \pi}\mbox{Re}[w_{l',l}(\Phi) \times \nonumber \\
& &  G^R_{l,l_{c \alpha},\sigma}(\omega) \Sigma^<_{eff, \alpha \beta}(\omega) 
G^A_{l_{c \beta},l',\sigma}(\omega)],
\end{eqnarray}
for the current along a given bond $\langle l, l' \rangle $ and
\begin{eqnarray}\label{jal}
J_{\alpha}& = &  - 2  \sum_{\sigma,\alpha=1}^M \int_{-\infty}^{+\infty}
\frac{ d\omega}{2 \pi} \mbox{Re}[\Sigma^<_{eff, \alpha \beta}(\omega) 
G^A_{l_{c \beta},l_{c \alpha},\sigma}(\omega) \nonumber \\
& &  
+  \Sigma^R_{eff, \alpha \beta}(\omega) 
G^<_{l_{c \beta},l_{c \alpha},\sigma}(\omega)],
\end{eqnarray}
for the current along the contact to the wire $\alpha$.
Details for the derivation of the latter equation from (\ref{curwi}) follow the same lines
as in Refs. \onlinecite{meiwi1,meiwi2} 
(see e.g. Eq. (5) of Ref. \onlinecite{meiwi1}), using
the normal Green's functions (\ref{gret}) and (\ref{gless}).

\subsection{Equilibrium currents.}
When the central system is attached to wires at the same chemical potential $\mu$, there is no charge flow
through the contacts to the reservoirs. Nevertheless,  if the central system is threaded by a finite magnetic
flux, equilibrium currents can flow within this system. For a given bond $\langle l, l'\rangle$, the
equilibrium current reads:
\begin{eqnarray} \label{teq}
J_{l,l'}^{eq} &= & \int_{-\infty}^{+\infty} \frac{d\omega}{2\pi}
f(\omega) T_{l,l'}^{eq}(\omega), \nonumber\\
T_{l,l'}^{eq}(\omega)&= & - 2  \mbox{Re} \{ w_{l',l}(\Phi)[ G^A_{l,l',\sigma}(\omega)-G^R_{l,l',\sigma}(\omega) ] \}\nonumber\\
& = & 
2  \mbox{Im}[\sum_{\sigma,\alpha,\beta=1}^M \Gamma_{eff,\alpha \beta}(\omega) w_{l',l}(\Phi) \nonumber\\
& & \times G^R_{l,l_{c \alpha},\sigma}(\omega)G^A_{l_{c \beta},l',\sigma}(\omega)],
\end{eqnarray}
where we have used the equilibrium identities
 (\ref{sieq}) and (\ref{geq}), while
 $\Gamma_{eff,\alpha \beta}(\omega)$ is defined in Eq. (\ref{gamaef}).
 For $\Phi=0$, the result $T_{l,l'}^{eq}(\omega)|_{\Phi=0}=0$ is obtained by noticing that 
the 
 function within
$[\ldots ]$ of the  above expression 
is just the real function
 $-2 \mbox{Im}[w_{l',l}(0)G^R_{l,l'}(\omega)|_{\Phi=0}]$.

\subsection{Non-equilibrium currents.}
We consider $M_{\cal S}$ S-wires at $\mu_{\alpha} \equiv \mu$ 
and $M_{\cal N}=M-M_{\cal S}$ N-wires with a voltage difference $V$ with respect to
the superconducting ones. 
Following Ref. \onlinecite{cue} we take $\mu_{\alpha} \equiv \mu$ in the Hamiltonians
$H_{\alpha}$ for the N-wires and enclose the bias $V$ in the corresponding
Fermi functions. We also consider that all the wires are at the same
temperature. Therefore, for the N-wires: 
$\Sigma^{gg,<}_{\alpha}(\omega)=  i f(\omega-V) \Gamma_{\alpha}(\omega)$ 
and
$\Sigma^{ff,<}_{\alpha}(\omega)=  i f(\omega+V) \Gamma_{\alpha}(\omega)$,
where $ \Gamma_{\alpha}(\omega) \equiv \Gamma^{gg}_{\alpha}(\omega)|_{\Delta_{\alpha}=0}$,
while for the superconducting ones: 
$\Sigma^{\nu \nu',<}_{\alpha}(\omega)=  i f(\omega) \Gamma^{\nu \nu'}_{\alpha}(\omega)$,
with $\nu, \nu'=g,f$.
In order to derive the expressions for the currents it is useful to express
the effective lesser self-energy as follows:
\begin{eqnarray}
\Sigma^<_{eff,\alpha,\beta}(\omega) & &= i f(\omega) \Gamma_{eff,\alpha,\beta}(\omega) +  \\
& &  i [f(\omega - V) - f(\omega)] \delta_{\alpha,\beta} \sum_{\alpha' \in {\cal N}} 
 \delta_{\alpha,\alpha'} \Gamma_{\alpha'}^{gg}(\omega) + \nonumber \\
& & i [f(\omega + V) - f(\omega)] \sum_{\alpha' \in {\cal N}} \Lambda^R_{\alpha, \alpha'}
 \Gamma_{\alpha'}^{ff}(\omega)\Lambda^A_{\alpha', \beta},\nonumber
\end{eqnarray}
where $\Gamma_{eff,\alpha,\beta}(\omega)$ has been defined in Eq. (\ref{gamaef}).

The final expression for the non-equilibrium current along a given bond of nearest neighbors $\langle l, l' \rangle $ is:
\begin{eqnarray}\label{curr}
J_{l,l'}& &=\int_{-\infty}^{+\infty}
 \frac{ d\omega}{2 \pi} [f(\omega-V)-f(\omega)] T(\omega).
\end{eqnarray}
In the case that, in addition to the bias $V$, the central system is threaded by a magnetic
flux, we should add to the previous expression the equilibrium contribution
$ J^{eq}_{l,l'}$ defined in the previous subsection. $ J^{eq}_{l,l'}$ 
is due to the internal currents of the single-electron  orbits of the finite system 
that are twisted by
the static flux. Instead, the origin of the non-equilibrium contribution is a net particle
 flow between
reservoirs through the central structure. For this reason, the non-equilibrium component depends only
on the spectral properties within the energy window $[\mu,\mu+V]$, while the equilibrium
one formally depends on the spectral weight of 
all the quantum states bellow $\mu$.

The  transmission function contains two contributions:
\begin{equation}
T_{l,l'}(\omega)=T^n_{l,l'}(\omega) - T^a_{l,l'}(-\omega).
\end{equation}
The first one is the 
normal transmission function:
\begin{eqnarray}
T^n_{l,l'}(\omega) & & = 2 \sum_{\sigma,\alpha \in {\cal N}=1}^{M_{\cal N}}
 \Gamma_{\alpha}^{gg}(\omega)\times \nonumber \\
& &
 \mbox{Im}[w_{l',l}(\Phi) G^R_{l,l_{c \alpha},\sigma}(\omega)  G^A_{l_{c \alpha},l',\sigma}(\omega) ],
\end{eqnarray}
and the second one is the Andreev transmission function,
\begin{eqnarray}
T^a_{l,l'}(\omega)& & =  - 2 \sum_{\sigma,\alpha \in {\cal N}=1}^{M_{\cal N}}\Gamma_{\alpha}^{ff}(\omega) \times 
\nonumber \\
& &
\mbox{Im}[w_{l',l}(\Phi) \overline{\Lambda}^R_{l,\alpha,\sigma}(\omega)  
\overline{\Lambda}^A_{\alpha,l',\sigma}(\omega) ],
\end{eqnarray}
where the  $\alpha \in {\cal N}$ denotes summation over the normal wires, 
while $\overline{\Lambda}^R_{l,\alpha,\sigma}(\omega) =
\sum_{\beta} G^R_{l,l_{c \beta},\sigma} \Lambda^R_{\beta,\alpha}(\omega)$ and 
 $\overline{\Lambda}^A_{\alpha,l'}(\omega) = 
[\overline{\Lambda}^R_{l',\alpha}(\omega)]^*$.
While the normal transmission function depends on the rate at which electrons can be
emitted at the normal reservoirs $ \Gamma_{\alpha}^{gg}(\omega)$, the Andreev
transmission function depends on the rate of emission of  holes (we recall that 
$\Gamma_{\alpha}^{ff}(\omega)=\Gamma_{\alpha}^{gg}(-\omega)$). The Andreev component depends on
the multiple scattering propagators  
$\overline{\Lambda}^R_{l,\alpha,\sigma}(\omega)$. Instead, the normal
component  depends on the usual ones $ G^R_{l,l_{c\alpha},\sigma}(\omega)$.  
For a vanishing superconducting gap, $T^a_{l,l'}(\omega) = 0$, and only the normal component survives.

Analogously, the currents through the contacts can be written as:
\begin{eqnarray}
J_{\alpha}&=& \int_{-\infty}^{+\infty}
\frac{ d\omega}{2 \pi} [f(\omega-V)-f(\omega)]
T_{\alpha}(\omega),
\end{eqnarray}
with the transmission function also containing two components:
\begin{eqnarray}
T_{\alpha}(\omega) & = & T^n_{\alpha}(\omega)- T^a_{\alpha}(-\omega).
\end{eqnarray}
The normal transmission function reads:
\begin{eqnarray} \label{tnal}
 T^n_{\alpha}(\omega) &=& 2 \sum_{\sigma,\beta =1}^{M} \{
\delta_{\alpha,\beta} \Gamma_{\alpha}^{gg}(\omega) 
\mbox{Im}[G^A_{l_{c \alpha} l_{c \alpha},\sigma}(\omega)]+ \nonumber \\
& & \sum_{\alpha' \in {\cal N}=1}^{M_{\cal N}}\Gamma_{\alpha'}^{gg}(\omega) 
\mbox{Im}[\Sigma^R_{\alpha \beta}(\omega) \nonumber \\
& & \times
G^R_{l_{c \beta} l_{c \alpha'},\sigma}(\omega)
G^A_{l_{c \alpha'} l_{c \alpha},\sigma}(\omega)] \},
\end{eqnarray}
while the Andreev transmission function is:
\begin{eqnarray}
 T^a_{\alpha}(\omega) &=& - 2 \sum_{\sigma,\beta =1}^{M} \sum_{\alpha' \in {\cal N}=1}^{M_{\cal N}}
\Gamma_{\alpha'}^{ff}(\omega) \nonumber \\
& & \times
\mbox{Im}[\Lambda^R_{\alpha, \alpha',\sigma}(\omega)
\Lambda^A_{\alpha', \beta,\sigma}(\omega) G^A_{l_{c \beta} l_{c \alpha},\sigma}(\omega)+
\nonumber \\
& & 
\Sigma^R_{eff, \alpha \beta}(\omega) \overline{\Lambda}^R_{l_{c \beta}, \alpha',\sigma}(\omega)
\overline{\Lambda}^A_{\alpha', l_{c \alpha},\sigma}(\omega)].
\end{eqnarray}

\section{A linear biased setup with a single superconducting wire and a single normal wire.}
In this section, we shall explicitly write down the previous expressions for the case
of a setup with two wires: one superconducting and the other one normal, which
we denote, respectively, $\alpha=N$ and $\alpha=S$. This will
allow us to show that we are able to recover
BTK's description \cite{btk} for the transmission functions of a simple
  tunneling junction. 

In this case: $\Sigma^R_{eff, \alpha \beta}(\omega) = 
\delta_{\alpha, \beta} [\delta_{\alpha,N}
\Sigma^{gg,R}_{N}(\omega) + \delta_{\alpha,S}
\Sigma^{gf,R}_{S}(\omega)\overline{g}^R_{l_S,l_S}(\omega)\Sigma^{fg,R}_{S}(\omega)$.
The total transmission function evaluated at the contact with 
the $N$-wire is $T_N(\omega)=T_N^n(\omega)-T_N^a(-\omega)$. The normal component is given
by Eq. (\ref{tnal}), which in this simple case reduces to:
\begin{eqnarray}
T^n_N(\omega)=\sum_{\sigma} \Gamma^{gg}_N(\omega) 
|G^R_{l_N,l_S, \sigma}(\omega)|^2 \Gamma^{gg}_{eff,S}(\omega).
\end{eqnarray}
Notice that we recover the well known structure for the
normal transmission function in terms of Green's functions originally pointed out  by 
Fisher and Lee \cite{file}. In the present case,
the function  $\Gamma^{gg}_{eff,S}(\omega)=\Gamma^{gg}_{S}(\omega)-
2 \mbox{Im}[\Sigma^{gf,R}_{S}(\omega)\overline{g}^R_{l_S,l_S}(\omega)
\Sigma^{fg,R}_{S}(\omega)]$ contains the usual term $\Gamma^{gg}_{S}(\omega)$, which
depends on the normal density of states of the superconducting lead,
as well as a multiple-scattering term that depends on the hole propagator 
$\overline{g}^R_{l_S,l_S}(\omega)$ and the anomalous self-energy of the wire
$\Sigma^{gf,R}_{S}(\omega)$.
The Andreev transmission function reads:
\begin{equation}
T^a_N(\omega)= - \sum_{\sigma} \Gamma^{ff}_N(\omega) 
|\overline{\Lambda}^R_{N,N,\sigma}(\omega)|^2 \Gamma^{gg}_{N}(\omega),
\end{equation}
which actually has the formal structure of a reflection process
represented in terms of Green's functions. Furthermore, it
 depends on the emission rate for holes in the normal wire
$\Gamma^{ff}_N(\omega)=\Gamma^{gg}_N(-\omega)$ and it contains
a multiple scattering kernel:
\begin{eqnarray}
\overline{\Lambda}^R_{N,N,\sigma}(\omega) & = &
G^R_{l_N,l_S,\sigma}(\omega) \Lambda^R_{S,N}(\omega),\nonumber \\
\Lambda^R_{S,N}(\omega) & = & \Sigma^{gf,R}_{S}(\omega) \overline{g}^R_{l_S,l_N}(\omega).
\end{eqnarray}
After some algebra, it can be verified that: 
$T^a_N(\omega)=-T^a_S(\omega) \equiv T^a(\omega) $
and $T^n_N(\omega)=-T^n_S(\omega)=T^n(\omega)$, in consistency with the continuity of the current.

In order to benchmark  the above representation,  we present results for the central system being
 a linear one-dimensional
junction with a barrier of height $E_0$
as in BTK's paper \cite{btk} (see also Ref. \onlinecite{aflec}):
\begin{eqnarray} \label{linear}
H_{cen} & = &-w \sum_{l=-1}^{0} 
\sum_{\sigma} [c^{\dagger}_{l, \sigma} c_{l+1,\sigma} + H.c] \nonumber \\
& &
+ \sum_{l=-1}^1 \varepsilon^0_l n_l,
\end{eqnarray}
with $ n_l=\sum_{\sigma} c^{\dagger}_{l, \sigma} c_{l,\sigma}$
and $\varepsilon^0_l= -\mu + \delta_{l,0}E_0$.
For such a system, it is easy to verify that the expressions for the transmission functions
corresponding to a given bond $\langle l, l+1 \rangle $ 
are:
\begin{eqnarray}
T^n_{l,l+1}(\omega)& = & 2 w \sum_{\sigma}
\mbox{Im}[G^R_{l,l_N,\sigma}(\omega)G^A_{l_N,l+1,\sigma}(\omega)]
\Gamma^{gg}_N(\omega),\nonumber \\
T^a_{l,l+1}(\omega)& = & - 2 w \sum_{\sigma}\mbox{Im}[G^R_{l,l_S,\sigma}(\omega)
G^A_{l_S,l+1,\sigma}(\omega)]\nonumber \\
& & \times |\Lambda_{S,N,\sigma}(\omega)|^2
\Gamma^{ff}_N(\omega).
\end{eqnarray}
It can be proved that this functions also satisfy $T^n_{l,l+1}(\omega)=T^n(\omega)$ and 
$T^a_{l,l+1}(\omega)=T^n(\omega)$, in agreement with the conservation of the current.

\begin{figure}
\includegraphics[width=0.9\columnwidth,clip]{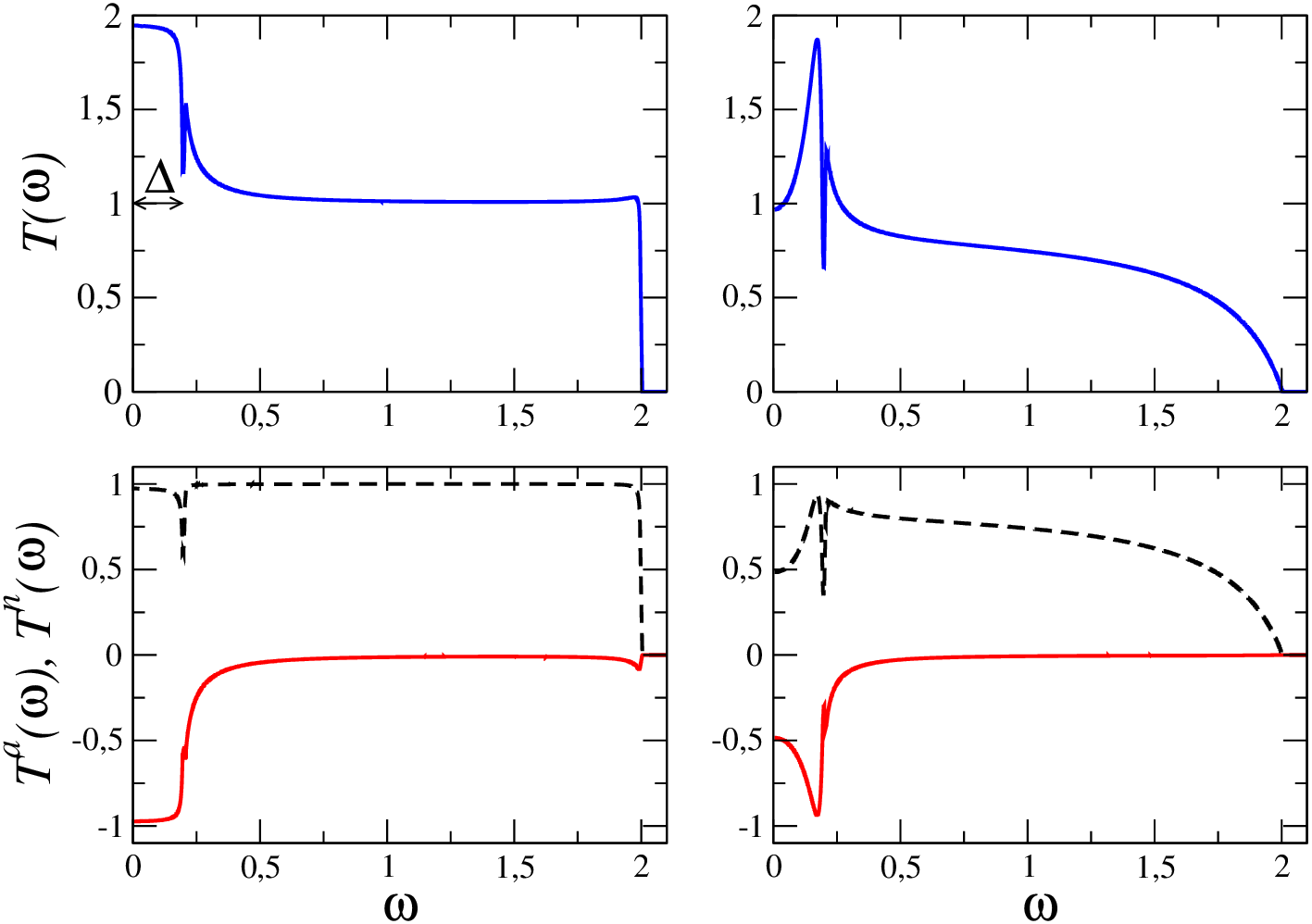}
\caption{\label{fig2} (Color online) Benchmark against BTK theory.
 Transmission functions $T^n(\omega)$ (dashed black lines) and $T^a(\omega)$ (red solid lines)
in the
lower panels and the total transmission $T(\omega)=T^n(\omega)-T^a(\omega)$ in the
upper panels for a junction described by the Hamiltonian (\ref{linear}).
Left and right panels correspond to $E_0=0, 1$, respectively. Other
parameters are $w_N=w_S=w=1$, $\mu=0$ and $\Delta_S=0.2$.}
\end{figure}

Numerical results for the functions
$T^n(\omega)$ and $T^a(\omega)$ are shown in the lower panels of
 Fig. \ref{fig2}. The corresponding total
transmission $T(\omega)$
 is also shown in the upper panels for $E_0=0$ and $E_0=1$.
 The picture presented in BTK's paper \cite{btk} is identified through 
  $T^n(\omega) \rightarrow 1-B(E)$ and 
$T^a(\omega) \rightarrow -A(E)$, with $A(E), B(E)$ defined in
Ref. \onlinecite{btk}. The lower panels of Fig. 2 should be compared with
Fig.5 of Ref. \onlinecite{btk}.
 It is worth noticing, in particular, the fact that 
$T^a(\omega)$ is sizable within the gap, while
 in the absence of a barrier ($E_0=0$),
 $T^a(\omega)\rightarrow -1$. Thus
$T(\omega) \sim 2$ for $|\omega|\leq \Delta$, (see upper panels of Fig. 2 and 
compare with Fig.7 of Ref.\onlinecite{btk}). 

\section{Flux sensitivity of the equilibrium currents in a ring.}
\begin{figure}
\includegraphics[width=0.9\columnwidth,clip]{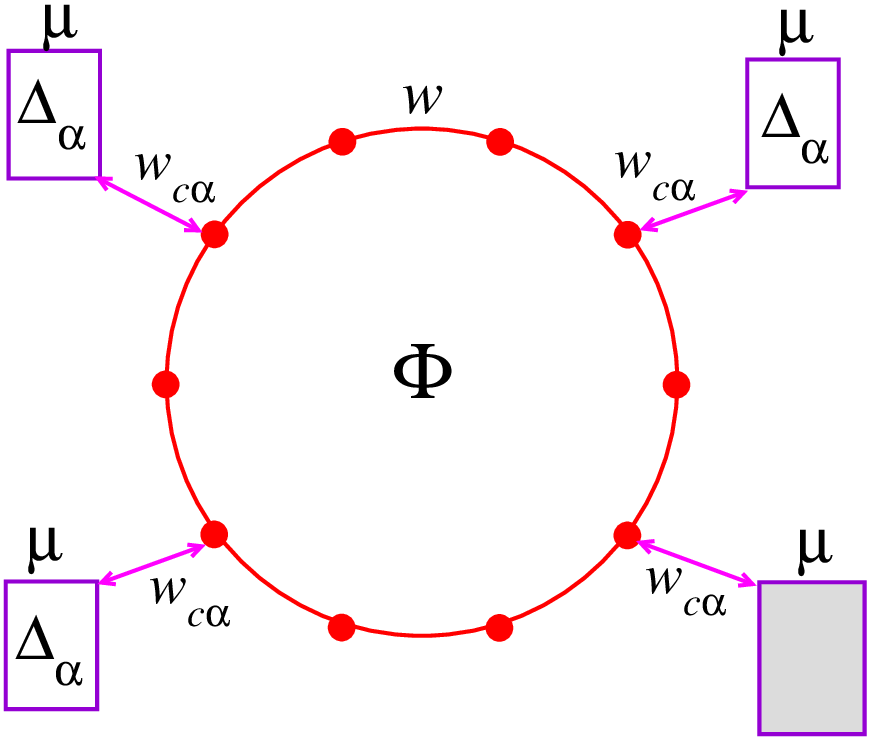}
\caption{\label{fig3} (Color online) Sketch of the setup. The central 
system is a ring threaded by a magnetic flux in contact to superconducting
and normal reservoirs at the same chemical potential $\mu$. The only non-vanishing
current is the equilibrium current along the circumference of the ring.}
\end{figure}
We now turn to the setup without bias voltage
($V=0$). We consider the simple case sketched in Fig. \ref{fig3}, where the central
system corresponds  to
  a one-dimensional  ring threaded by a magnetic flux $\Phi$, 
i.e. $H_{cen} \equiv H_{ring}$, being:
\begin{eqnarray}
H_{ring}& = & -w \sum_{l=1,\sigma}^L (e^{- i \Phi/L} 
c^{\dagger}_{l, \sigma} c_{l+1, \sigma} + H.c.) \nonumber \\
& & +\sum_{l=1,\sigma}^L \varepsilon^0_l c^{\dagger}_{l, \sigma} c_{l, \sigma} ,
\end{eqnarray}
where $\Phi$ is expressed in units of $2 \pi \Phi_0$, being  
$\Phi_0=e/h$ the elementary quantum.
We take the lattice constant $a=1$ and we impose the 
periodic boundary condition $L+1 \equiv 1$.

An isolated normal ring under a magnetic flux, supports a persistent current with a periodicity
equal to $\Phi_0$, as a consequence of the sensitivity of its energy levels 
with the threading flux. When normal metallic wires are attached to the ring, inelastic scattering
effects are introduced which  decrease  the magnitude of this equilibrium current. 
However,
its qualitative behavior, in particular, the periodicity with the flux is expected to be the same as
in the case of the isolated ring, provided that the inelastic scattering length $\xi_{in}$ introduced
by the coupling to the external wires satisfies $\xi_{in} > La $. For $\xi_{in} <La $, this current is, 
instead, expected to vanish. 
This is because, for a short enough ring such that $\xi_{in} > La $, the effect of the coupling to
the wires is essentially the introduce
tion of a finite lifetime  in the energy levels,
without affecting their flux sensitivity.

In the case of an isolated superconducting ring with s-wave pairing, Byers and Yang \cite{by} have shown 
that the periodicity of the flux-induced
persistent currents is $\Phi_0/2$. This is again a consequence of the sensitivity of the energy levels, this time combined
with the fact that the structure of the wave function corresponds to an ensemble of Cooper pairs, instead of one of
single electrons. Hybrid isolated S-N piecewise rings have been also studied and the conclusion is that the periodicity of the
persistent currents experiences a 
crossover between  $\Phi_0/2$ and  $\Phi_0$,
 as the length of the superconducting piece becomes shorter than the
superconducting coherence length $\xi_c$ \cite{ringbu,ringcay}.

On the other hand, a conductor between two superconductors forming a S-N-S structure is known to support Andreev states  within the 
superconducting gap. In particular, such states are expected to develop for a ring with attached superconducting wires and it is
interesting to study  the
flux sensitivity of these states, which should define the behavior of the equilibrium currents. It is also interesting 
to investigate which is the minimum number of S-wires needed to develop Andreev states.
 Furthermore, recent studies suggest that the vortex excitations of
a superconducting state can exist within a 
normal conductor sandwiched between two superconductors \cite{cueber} due to the proximity effect. It is, therefore 
interesting to investigate whether it is possible that proximity effect induces also  
a flux periodicity of $\Phi_0/2$ in a normal ring due to the attachment to S-wires.

In order to address these issues we analyze the behavior of the function 
$T^{eq}(\omega)$. Because of the continuity of the charge, this function is independent of the bond
 $l,l+1$ along the ring chosen for the evaluation of Eq. (\ref{teq}). Thus, 
the latter expression can also be written as follows: 
\begin{eqnarray}\label{teqring}
T^{eq}(\omega)& = &  -\frac{2 w}{L} \sum_{l=1}^L \sum_{\sigma, \alpha,\beta} \mbox{Re} 
\{ e^{-i \Phi/L} \nonumber \\
& & \times [ G^R_{l,l+1,\sigma}(\omega)
-[G^R_{l+1,l,\sigma}(\omega)]^* ] \}.
\end{eqnarray}
In what follows, we analyze different configurations of wires.

\subsection{Each site of the ring in contact with a wire.}
Let us first  consider the simple case of a ring in contact to wires in a configuration that does not break
the periodic translational invariance along the circumference of the ring. Such a configuration corresponds to
$L$ identical wires (N or S), each one in contact to a single site of the ring.
The retarded Green's function can be easily evaluated in this case. The 
result is:
\begin{eqnarray}
G^R_{l,l',\sigma}(\omega) & = & \frac{1}{L}
\sum_{m=0}^{L-1} e^{i k_m (l-l')} G^R_{m,\sigma}(\omega),\nonumber \\
 G^R_{m,\sigma}(\omega)& = &\frac{1}{\omega - \varepsilon_m(\Phi) - \Sigma_m^{eff,R}(\omega)}
\end{eqnarray}
with $k_m= -\pi +2 m \pi/L$, $m=0,\ldots,L-1$, and 
$\varepsilon_m(\Phi)= - 2 w \cos(k_m + \Phi/L)$, where, for simplicity, we have taken $\mu=0$.
The effective self-energy is:
\begin{equation}
\Sigma_m^{eff,R}(\omega)= \Sigma^{gg,R}(\omega)-
\Sigma^{gf,R}(\omega) \overline{g}^R_m(\omega) \Sigma^{fg,R}(\omega),
\end{equation}
where the second term vanishes for N-wires. The hole propagator of this term is:
\begin{equation}
 \overline{g}^R_m(\omega)=\frac{1}{\omega+ \varepsilon_m(-\Phi) - \Sigma^{ff,R}(\omega)}.
\end{equation}
Transforming the right hand side of (\ref{teqring}) to the reciprocal space, it reduces to:
\begin{equation}\label{teqrm}
T^{eq}(\omega)= \frac{2}{L} \sum_{m=0}^{L-1} v_m(\Phi) \{ -2 \mbox{Im}[G^R_m(\omega)] \},
\end{equation}
with $ v_m(\Phi) =2 w \sin(k_m -\Phi/L)=\partial \varepsilon_m(\Phi)/\partial k_m$ being the velocity corresponding to the $m$-th energy level.

In the limit where the coupling to the wires vanishes, the above expression reduces to the
transmission function of an isolated ring:
\begin{equation}
T^{eq}(\omega) \stackrel{w_{c \alpha} \rightarrow 0}{\longrightarrow} 
 \frac{4 \pi }{L} \sum_{m=0}^{L-1} v_m(\Phi) \delta(\omega - \varepsilon_m(\Phi)).
\end{equation}

For  N-wires or for S-wires and energies such that
$|\omega|> \Delta$, a similar expression is obtained:
\begin{equation}\label{lore}
T^{eq}(\omega) =  \frac{4 \Theta(|\omega|-\Delta)  }{L} \sum_{m=0}^{L-1}  
\frac{ v_m(\Phi) \mbox{Im}[\Sigma_m^{eff,R}(\omega)]}
{|\omega- \varepsilon_m(\Phi)- \Sigma_m^{eff,R}(\omega)|^2},
\end{equation}
where the $\Theta$-function applies only for the case of a S-wire. The above expression corresponds to a sequence of Lorenzian functions centered at energies
$\sim \varepsilon_m(\Phi) + \mbox{Re}[\Sigma_m^{eff,R}( \varepsilon_m(\Phi))]$
with width $\sim \mbox{Im}[\Sigma_m^{eff,R}( \varepsilon_m(\Phi))]$. The latter
parameter defines the lifetime of the levels of the ring due to the coupling
to the reservoirs.

The periodicity of these currents as functions of the flux is $\Phi_0$, which corresponds to
a shift $\Phi/L=2 \pi/L$, that is equivalent to a relabeling of the reciprocal points
$k_m$.
For S-wires and $|\omega|< \Delta$, the functions $\Gamma^{\nu,\nu'}(\omega)=0$, thus
$\mbox{Im}[\Sigma_m^{eff,R}(\omega)]=0$, and the only spectral contribution to $T^{eq}(\omega)$
is due to the eventual development of Andreev states. The energies of these states is determined
from the poles of the function $G^R_{m,\sigma}(\omega)$, which implies finding the
roots of the function:
\begin{eqnarray}
\lambda(\omega) & = &  \omega -\varepsilon_m(\Phi)- \mbox{Re}[\Sigma^{gg,R}(\omega)] \nonumber \\
& & - \mbox{Re}[\Sigma^{gf,R}(\omega) \Sigma^{fg,R}(\omega)] \mbox{Re}[\overline{g}^R_m(\omega)],
\end{eqnarray}
where
\begin{equation}
\overline{g}^R_m(\omega)= \Theta(\Delta-|\omega|) 
\frac{1}{\omega+ \overline{\varepsilon}_{-m}(\Phi) + i \eta},
\end{equation}
with $\overline{\varepsilon}_m(\Phi) \sim \varepsilon_m(\Phi) + 
\mbox{Re}[\Sigma^{gg}(\varepsilon_m(\Phi))]$.

Approximating $\mbox{Re}[\Sigma^{\nu,\nu'}(\omega)] \sim \mbox{Re}[\Sigma^{\nu,\nu'}
(\varepsilon_{\pm m}(\Phi))]$, the solution casts the following roots:
\begin{eqnarray}
& & E^{\pm}_m(\Phi) \sim e_m^-(\Phi)
 \pm \nonumber \\
& & \sqrt{ [e_m^+(\Phi) ]^2+
 \mbox{Re}[\Sigma^{gf}(\varepsilon_m(\Phi))\Sigma^{fg}(\varepsilon_m(\Phi))]},
\end{eqnarray}
with
\begin{equation}
 e_m^{\pm}(\Phi)=\frac{\overline{\varepsilon}_m(\Phi) \pm \overline{\varepsilon}_{-m}(\Phi)}{2}
\end{equation}
while the corresponding quasiparticle weights are:
\begin{equation}\label{zm}
Z^{\pm}_m = \frac{ -\pi}{|\partial \lambda(\omega)/\partial \omega |_{E_m^{\pm}} }\sim
 \frac{ -\pi}{2 |E_m^{\pm}|}.
\end{equation}
Replacing in (\ref{teqrm}), the final result for the transmission function within the superconducting gap is:
\begin{equation} \label{teqf}
T^{eq}(\omega) = 
 \frac{2 \pi \Theta(\Delta-|\omega|) }{L} \sum_{s=\pm,m=0}^{L-1} \frac{v_m(\Phi)}{|E_m^{\pm}(\Phi)|}
 \delta(\omega - E^s_m(\Phi)).
\end{equation}

For $|\omega|<\Delta$:
\begin{eqnarray}
\mbox{Re}[\Sigma^{gg}(\omega)]& = & \omega \gamma(\omega),\nonumber \\
\mbox{Re}[\Sigma^{gf}(\omega)]& = & \Delta \gamma(\omega),
\end{eqnarray}
being
\begin{eqnarray}
\gamma(\omega) & = & \frac{|w_{c}|^2}
{2 w_{\alpha}^2}[1-\sqrt{1+\frac{4 w_{\alpha}^2}{\Delta^2-\omega^2}}].
\end{eqnarray}
Therefore:
\begin{eqnarray}\label{epm}
E^{\pm}_m(\Phi)& & \sim
\beta [\varepsilon_m(\Phi)-\varepsilon_{-m}(\Phi)] \pm \nonumber \\
 & & \sqrt{\beta^2 [ \varepsilon_m(\Phi)+ \varepsilon_{-m}(\Phi)]^2+
\gamma^2 \Delta^2 },
\end{eqnarray}
being $\beta=(1+\gamma(\varepsilon_m(\Phi)))/2$, and  $\gamma \sim \gamma(\varepsilon_m(\Phi))$.

Remarkably, the expression (\ref{teqf}) with the energy given by (\ref{epm})
coincides with 
 the expression for the persistent currents of an 
isolated 1D BCS tight-binding ring  with hopping
$2 \beta w$, gap $2 \gamma \Delta$ and pairs with total momentum $q=0$
(see Ref. \onlinecite{lod}). In other words, the flux sensitivity of the Andreev states 
in our problem 
is exactly the same as that observed in an isolated BCS 1D ring with pairs of 
momentum $q=0$. The fact that only pairs with momentum $q=0$ contribute
 implies that the periodicity of these currents is just the normal periodicity
of a flux quantum $\Phi_0$. These currents do not show the $\Phi_0/2$ periodicity, 
typical of a true superconducting ring, since the origin of that behavior is a
change in $2 \pi/L$ of the total momentum $q$ of the Cooper pairs.
 The renormalization factor $\beta$ for the hopping parameter within the ring, which 
determines the velocity $v_m$ and, thus, the amplitude of the currents, 
depends on the 
superconducting coherence length of the wires, $\xi_c \sim \Delta/2 w$,
 as well as on the tunneling ratio through the contacts, controlled by the parameter $w_c$.
Its magnitude is large for energies close to the edge of the gap 
$|\varepsilon_m(\phi)| \sim \Delta$. 

\subsection{A single S-wire attached to the ring.}
Let us now consider a single superconducting wire attached to the ring.

As before, we must consider separately the contribution from states with energies within and
away from the superconducting gap. To analyze the spectrum for energies $|\omega|>\Delta$,
it is convenient to write the
retarded Green's function as follows:
\begin{equation}\label{ret1}
G^R_{l,l_{c \alpha},\sigma}(\omega)=\frac{g^{0}_{l,l_{c \alpha}}(\omega)}
{1-\Sigma^{R}_{eff,\alpha}(\omega)
g^{0}_{l_{c \alpha},l_{c \alpha}}(\omega)},
\end{equation}
being
\begin{eqnarray}\label{g0}
g^{0}_{l,l'}(\omega)&=&\frac{1}{L}\sum_{m=0}^{L-1} e^{-i k_m (l-l')} g^0_{k_m}(\omega),
\nonumber \\
g^{0}_{k_m}(\omega)&=&\frac{1}{\omega-\varepsilon_m(\Phi) + i \eta },
\end{eqnarray}
and $\Sigma^{R}_{eff,\alpha}(\omega)= \Sigma^{gg}_{\alpha}(\omega)+
 \Sigma^{gf}_{\alpha}(\omega) 
\overline{g}^0_{l_{c \alpha},l_{c \alpha}}(\omega)\Sigma^{fg}_{\alpha}(\omega) $, with:
\begin{eqnarray}\label{g0-}
\overline{g}^{0}_{l,l'}(\omega)&=&\frac{1}{L}\sum_{m=0}^{L-1} e^{-i k_m (l-l')} 
\overline{g}^0_{k_m}(\omega),
\nonumber \\
\overline{g}^{0}_{k_m}(\omega)&=&\frac{1}{\omega+\varepsilon_m(-\Phi) + i \eta }.
\end{eqnarray}

Substituting in (\ref{teqring}), the transmission function reads:
\begin{eqnarray}\label{teq1}
T^{eq}(\omega)& = &  \frac{2 \Theta(|\omega|-\Delta)}{L} \sum_{m=0}^{L-1}  v_m(\Phi) A_m(\omega),
\end{eqnarray}
being
\begin{equation}\label{an}
A_m(\omega)= \frac{\Gamma_{eff,\alpha,\alpha}(\omega)|g^0_{k_m}(\omega)|^2}
{|1-\Sigma^{R}_{eff,\alpha}(\omega)
g^{R}_{l_{c \alpha},l_{c \alpha}}(\omega)|^2},
\end{equation}
which results in a Lorentzian-type profile as in the case of Eq. (\ref{lore}).

As in the case considered in the previous subsection, for $|\omega|<\Delta$, 
$\mbox{Im}[\Sigma^{\nu \nu',R}_{\alpha}(\omega)]=0$, and Andreev states can 
develop within
the gap. In order to determine the energies of these levels, it is convenient to
consider the retarded Green's functions $g^R_{l,l'}(\omega)$ and 
$\overline{g}^R_{l,l'}(\omega)$, defined in Eqs. (\ref{overg}),
which in the present case are the solutions of the following Dyson's
equations:
\begin{eqnarray}
g^R_{l,l'}(\omega) & = & g^0_{l,l'}(\omega)+
g^R_{l,l_{c \alpha}}(\omega)\Sigma^{gg,R}_{\alpha}(\omega)
g^0_{l_{c \alpha},l'}(\omega), \nonumber \\
\overline{g}^R_{l,l'}(\omega) & = & \overline{g}^0_{l,l'}(\omega)+
\overline{g}^R_{l,l_{c \alpha}}(\omega)\Sigma^{ff,R}_{\alpha}(\omega)
\overline{g}^0_{l_{c \alpha},l'}(\omega).
\end{eqnarray}
Within the gap, these functions have, respectively, quasiparticle and 
quasihole
  states, behaving as follows:
\begin{eqnarray}
g^R_{l,l'}(\omega) & \sim & 
\frac{ \Theta(|\omega|-\Delta)}{L} 
\sum_{m=0}^{L-1} \frac{e^{-i k_m (l-l')} Z_n }
{\omega - \stackrel{\sim}{\varepsilon}_m (\Phi) + i \eta}, \nonumber \\
\overline{g}^R_{l,l'}(\omega) &  \sim & 
\frac{ \Theta(|\omega|-\Delta)}{L} 
\sum_{m=0}^{L-1}\frac{e^{-i k_m (l-l')} Z_{-n}  }
{\omega + \stackrel{\sim}{\varepsilon}_{-m} (\Phi) + i \eta},
\end{eqnarray}
being $\stackrel{\sim}{\varepsilon}_m (\Phi) \sim \varepsilon_m (\Phi)+ C
\mbox{Re}[\Sigma^{gg}_{\alpha}(\varepsilon_m (\Phi))]/L$, where $C=2$ for 
$\Phi= K \pi$ with $K$ integer while $C=1$ otherwise, and
$Z_m = -\pi 
\{ |1- C \partial \mbox{Re}[\Sigma^{gg,R}_{\alpha}(\omega)]/
\partial \omega |_{ \stackrel{\sim}{\varepsilon}_m (\Phi)} /L \}^{-1}$. In what follows,
we shall approximate $Z_m \sim -\pi $, which becomes exact in the limit $L \rightarrow
\infty$.

The full retarded Green's function is, in turn, determined from:
\begin{eqnarray}
G^{R}_{l,l',\sigma}(\omega)& = & g^R_{l,l'}(\omega)+
G^{R}_{l,l_{c \alpha},\sigma}(\omega) \Sigma^{gf,R}_{\alpha}(\omega)  \nonumber \\
& & \times 
\overline{g}^R_{l_{c \alpha},l_{c \alpha}}(\omega)
\Sigma^{fg,R}_{\alpha}(\omega)  g^R_{l_{c \alpha},l'}(\omega).
\end{eqnarray}
As in the previous section, the ensuing solution has a quasiparticle BCS-like structure:
\begin{eqnarray}
G^R_{l,l'}(\omega) & \sim & 
\frac{ \Theta(|\omega|-\Delta)}{L} 
\sum_{s=\pm,m=0}^{L-1} \frac{e^{-i k_m (l-l')} Z^{s}_m }
{\omega - E^s_m (\Phi) + i \eta},
\end{eqnarray}
with $E^{\pm}_m (\Phi)$ given in (\ref{epm}), with $\gamma \propto 1/L$ and $Z^{\pm}_m$
given in Eq. (\ref{zm}).

Therefore, for a single superconducting wire connected to a large enough ring,
Andreev levels tend to coincide with
free particle and hole energies: $\varepsilon_m(\Phi)$ and $-\varepsilon_{-m}(\Phi)$,
respectively,
provided that $|\varepsilon_m(\Phi)|<\Delta$,  $|\varepsilon_{-m}(\Phi)|<\Delta$.
The corresponding transmission function is formally given by Eq. (\ref{teqf}).

In conclusion, a single superconducting wire attached to the ring generates the same
qualitative behavior as L superconducting wires attached in a translational symmetrical
way, but the effect is ${\cal O}(1/L)$ and tends to be negligible as $L \rightarrow \infty$.

\section{Summary and conclusions.}
We have presented a representation of Keldysh Green's functions for stationary 
transport problems in systems with superconducting and normal components. As most
of the relevant observables, like the currents, depend on normal propagators, we
have worked with Dyson's equations in order to eliminate the anomalous ones. This
procedure has been carried out by
defining auxiliary hole propagators and effective self-energies that contain multiscattering
terms. 
In the resulting representation, the Green's functions
 exhibit the same structure as in  normal systems. This 
allows for the derivation of simple and compact expressions for the
currents and the transmission functions,  that are similar to the ones
presented in Refs. \onlinecite{meiwi1,meiwi2} for 
normal systems. 

We have presented general expressions for the currents in stationary conditions, distinguishing
two situations: biased systems where transport is induced by a voltage difference and 
equilibrium currents induced by a static magnetic flux. In the case of biased systems, 
we have defined normal and Andreev transmission functions
and we have compared them with results obtained in the framework of previous formalisms,
in particular, the one presented by Blonder, Tinkham and Klapwijk.

We have, finally focused in the study of
the behavior of the equilibrium currents in a tight-binding 
normal ring with attached
superconducting wires. These currents result as superpositions of the currents of all the 
states of the ring with energies $\varepsilon_m(\Phi)$ bellow  the chemical potential of the wires, in which
electrons circulate with velocities $v_m=\partial \varepsilon_m(\Phi)/\partial k_m$.

Our main conclusions on the qualitative behavior of these currents are the following:
(i) The states with energies lying away from the energy window defined 
by the superconducting gap present an identical qualitative behavior as those of rings
attached to N wires. In particular, they have a periodicity of $\Phi_0$ as functions of the
external flux. The spectral profile related to these currents is a collection of Lorentzian
functions which implies a decrease in the amplitude of the current due to inelastic scattering
effects via the escape to the leads. 

  (ii) The states with energies within the superconducting gap of the wires, behave as isolated
in the sense that the spectral weight related to them consists in a collection of delta functions,
indicating the lack of inelastic scattering effects. The positions of the energy levels is, 
however, affected by the proximity effect and they are organized in a structure that replicates
the quasiparticle spectrum of a BCS tight-binding
 superconducting ring with Cooper pairs of momentum $q=0$. 
The effective BCS tight-binding parameters are the hopping, which is the bare hopping
of the ring renormalized by a factor $\beta$ and a gap, which is the gap of the superconducting wires
renormalized by a factor $\gamma$. The renormalizing factors 
depend on the superconducting
coherence length of the wires and the degree of coupling between the wires and the ring. The latter
effect is controlled by the strength of the coupling between these systems as well as on  the
number of attached wires. For a single attached wire, it is ${\cal O}(1/L)$ and, thus, not 
significant for large enough rings. 

(iii) Although the proximity effect induces Andreev levels that replicate
the structure of quasiparticle states of a superconducting ring within the energy window defined
by the superconducting gap of the wires, these states correspond only to the subspace with winding
number $q=0$. Since the periodicity in $\Phi_0/2$ 
of the persistent currents in superconducting rings is explained by a shift in the winding number
$q$ commensurate with the reciprocal lattice of the ring, \cite{by,lod}
 the restriction of the subspace with
$q=0$ does not allow for such a mechanism. The consequence of this rigidity is that Andreev
states have the same periodicity $\Phi_0$ as the states of the normal ring. Let us, however, mention
that the rigidity of the winding number could be due to the rigid BCS mean field approximation
considered to model the external wires. There exists the possibility that a more flexible model
allowing for spacial fluctuations of the parameter $\Delta$ within a region of the external wires
that is close to the contacts could also permit fluctuations in the winding number $q$ of the induced
Andreev sates within the ring. A possibility to explore this mechanism is by recourse to a
self-consistent approximation similar to that of Refs. \onlinecite{alf} and \onlinecite{sols}.

\section{Acknowledgements.}
The author thanks A. Aligia, M. B\"uttiker, H. Bouchiat and G. Lozano for useful  
comments and references. Support from CONICET and UBACYT Argentina, and from  
the ``RyC'' program from MCEyC of Spain is acknowledged.

\appendix

\section{Eliminating the degrees of freedom of the reservoirs.}\label{appintout}
We summarize the procedure introduced in Ref. \onlinecite{past,meiwi1,meiwi2} to eliminate the degrees
of freedom of the external wires in the Dyson's equation for the central system.

It is convenient to change the basis in $H_{\alpha}$ as follows:
\begin{equation}
c_{j_{\alpha}, \sigma} = \sqrt{\frac{2}{N_{\alpha}+1}} \sum_{n=0 }^{N_{\alpha}} 
\sin( k_{n,\alpha} j_{\alpha}) c_{k_{n, \alpha} \sigma},
\end{equation}
with $k_{n, \alpha}= n \pi/(N_{\alpha}+1), n=0, \ldots, N_{\alpha}$,
which leads to:
\begin{eqnarray}
H_{\alpha} & = &  \sum_{n=0}^{N_{\alpha}}\sum_{\sigma}
[ \varepsilon_{k_{n,\alpha}} c^{\dagger}_{k_{n,\alpha}, \sigma} c_{k_{n,\alpha},\sigma} \nonumber \\
& & +
\sum_{n=0}^{N_{\alpha}} \Delta_{\alpha}  c^{\dagger}_{k_{n,\alpha}, \uparrow}  
c^{\dagger}_{k_{n,\alpha}, \downarrow} +H.c],
\end{eqnarray}
being  $\varepsilon_{k_{n,\alpha}} = -2 w_{\alpha} \cos k_{n,\alpha} - \mu_{\alpha}$, and
\begin{equation}
H_{c,\alpha}=\sum_{n=0}^{N_{\alpha}}\sum_{\sigma} 
w_{\alpha,k} (c^{\dagger}_{k_{n,\alpha},\sigma} c_{l_{c \alpha},\sigma} + H.c),
\end{equation}
being $w_{\alpha,k}= -  \sqrt{\frac{2}{N_{\alpha}+1}} \sin k_{n, \alpha} w_{c \alpha} $.

Let us focus in the Dyson's equation  with coordinates $l_{c \alpha}, l'$,  belonging to the central system:
\begin{eqnarray}\label{gcal}
& & \omega G^R_{l_{c,\alpha},l',\sigma}(\omega) -  \sum_n w_{\alpha,k} G^R_{k_{n,\alpha},l',\sigma}(\omega) \nonumber \\ 
& & - \sum_{l''} 
\varepsilon_{l_{c, \alpha},l''}  G^R_{l'',l',\sigma}(\omega) = \delta_{l_{c,\alpha},l'}, \nonumber \\ 
& & \omega F^R_{l_{c,\alpha},l',\sigma}(\omega) +  \sum_n w_{\alpha,k} F^R_{k_{n,\alpha},l',\sigma}(\omega) \nonumber \\ 
& & + \sum_{l''} 
\varepsilon_{l_{c, \alpha},l''}  F^R_{l'',l',\sigma}(\omega) = 0, 
\end{eqnarray}
where $l''$ runs over all the spacial indexes of the central system while $k_{n,\alpha}$ labels degrees of freedom
of the reservoir represented by $H_{\alpha}$. The Green's functions with mixed coordinates $k_{n,\alpha},l'$, in turn,
satisfies the following equation:
 \begin{eqnarray}
& & \omega G^R_{k_{n,\alpha},l',\sigma}(\omega) - \varepsilon_{k_{n,\alpha}} G^R_{k_{n,\alpha},l',\sigma}(\omega) \nonumber \\
& & -w_{\alpha,k} G^R_{l_{c,\alpha},l',\sigma}(\omega) - \Delta_{\alpha} F^R_{k_{n,\alpha},l',\sigma}(\omega)=0,
\nonumber \\
& & \omega F^R_{k_{n,\alpha},l',\sigma}(\omega) + \varepsilon_{k_{n,\alpha}} F^R_{k_{n,\alpha},l',\sigma}(\omega) \nonumber \\
& & + w_{\alpha,k} F^R_{l_{c,\alpha},l',\sigma}(\omega) - \Delta^*_{\alpha} G^R_{k_{n,\alpha},l',\sigma}(\omega)=0.
\end{eqnarray}
After some algebra, the  above equations can be casted as follows:
\begin{eqnarray}\label{fgbond}
F^R_{k_{n,\alpha},l',\sigma}(\omega) & = & \overline{g}^{R,0}_{k_{n,\alpha}}(\omega)[\Delta^*_{\alpha} G^R_{k_{n,\alpha},l',\sigma}(\omega) \nonumber \\
& & - w_{\alpha,k} F^R_{l_{c,\alpha},l',\sigma}(\omega)], \\
G^R_{k_{n,\alpha},l',\sigma}(\omega) & = & w_{\alpha,k} 
[G^{R,0}_{k_{n,\alpha}}(\omega) G^R_{l_{c,\alpha},l',\sigma}(\omega)\nonumber \\
& & + F^{R,0}_{k_{n,\alpha}}(\omega) F^R_{l_{c,\alpha},l',\sigma}(\omega)] ,
\end{eqnarray}
with:
\begin{eqnarray}
\overline{g}^{R,0}_{k_{n,\alpha}}((\omega) &=& \frac{1}{\omega + \varepsilon_{k_{n,\alpha}} + i \eta}, \nonumber \\
G^{R,0}_{k_{n,\alpha}}((\omega) &=& \frac{(\omega + \varepsilon_{k{n,\alpha}})}{(\omega + i \eta)^2- E^2(\varepsilon_{k_{n,\alpha}})}, \nonumber \\
F^{R,0}_{k_{n,\alpha}}((\omega) &=& \frac{\Delta_{\alpha}}{(\omega + i \eta)^2- E^2(\varepsilon_{k_{n,\alpha}})},
\end{eqnarray}
with $\eta = 0^+$ and $ E^2(\varepsilon_{k_{n,\alpha}})= \varepsilon_{k_{n,\alpha}}^2+ \Delta_{\alpha}^2$.

Substituting  (\ref{fgbond}) into (\ref{gcal}), the latter equations can be expressed in the following way:
\begin{eqnarray}
& & [\omega  -  \Sigma^{gg,R}_{\alpha}(\omega)]G^R_{l_{c,\alpha},l',\sigma}(\omega)  +  \Sigma^{gf,R}_{\alpha}(\omega) F^R_{l_{c,\alpha},l',\sigma}(\omega) \nonumber \\ 
& & - \sum_{l''} 
\varepsilon_{l_{c, \alpha},l''}  G^R_{l'',l',\sigma}(\omega) = \delta_{l_{c,\alpha},l'}, \nonumber \\ 
& & [ \omega - \Sigma^{ff,R}_{\alpha}(\omega)] F^R_{l_{c,\alpha},l',\sigma}(\omega)  +  \Sigma^{fg,R}_{\alpha}(\omega) G^R_{l_{c,\alpha},l',\sigma}(\omega)\nonumber \\ 
& &
+  \sum_{l''} 
\varepsilon_{l_{c, \alpha},l''}  F^R_{l'',l',\sigma}(\omega) = 0.
\end{eqnarray}
Notice that all the spacial indexes of the above equations run over coordinates of the central system, while the
indexes corresponding to the reservoirs have been eliminated by defining the  
 `self-energies':
\begin{eqnarray}\label{self}
\Sigma^{\nu \nu',R}_{\alpha}(\omega) &=& \sum_n |w_{\alpha,k}|^2 \frac{\lambda^{\nu,\nu'}(\omega,\varepsilon_{k_{n,\alpha}})}
{\omega^2 - E(\varepsilon_{k_{n,\alpha}})^2},
\end{eqnarray}
being  $\lambda^{\nu, \nu'} (\omega, \varepsilon_{k_{n,\alpha}})= \delta_{\nu,\nu'} (\omega \pm \varepsilon_{k_{n,\alpha}})$ for 
$\nu=g,f$, respectively
and $\lambda^{g,f}(\omega, \varepsilon_{k_{n,\alpha}})= [\lambda^{f,g}(\omega, \varepsilon_{k_{n,\alpha}})]^*=\Delta_{\alpha}$.

These steps can be repeated with each contact, which allows for the one by one elimination of the degrees of freedom of all the wires.
The limit to the size of the wires going to infinite is summarized in Appendix \ref{appendixa}.

\section{Retarded self-energies associated a 1D S-wire.}\label{appendixa}
We now evaluate the spectral functions $\Gamma^{\nu, \nu'}_{\alpha}(\omega)
= -2 \mbox{Im}[\Sigma^{\nu \nu',R}_{\alpha}(\omega)]$ corresponding to the self-energies defined in the previous appendix
in the  thermodynamic limit, $N_{\alpha} \rightarrow \infty$. This corresponds to replacing  $\sum_n \rightarrow (N_{\alpha}/ \pi)
 \int_0^{\pi} dk$ in the expressions (\ref{self}):
\begin{eqnarray}\label{gamas}
\Gamma^{\nu  \nu'}_{\alpha}(\omega)& = &
 \frac{|w_{c \alpha}|^2}{2 w_{\alpha}^2}
\int_{-2 w_{\alpha}-\mu}^{2 w_{\alpha}-\mu} 
d u \lambda^{\nu, \nu'} (\omega, u)\nonumber \\
& & \times \frac{\sqrt{(2w_{\alpha})^2 -(u+\mu)^2} }{E(u)}
\nonumber \\
& & \times  \{ \delta(\omega - E(u))-\delta(\omega + E(u))\}.
\end{eqnarray}
The final result is:
\begin{eqnarray}
\Gamma^{gg}_{\alpha}(\omega)&=&
\Gamma^{ff}_{\alpha}(-\omega) = \mbox{sg}(\omega) \frac{|w_{c, \alpha}|^2}{2 w_{\alpha}^2}
\frac{1}{r(\omega)}\nonumber \\
& & \times \{[\omega+ r(\omega)]s^+(\omega)+
[\omega- r(\omega)]s^-(\omega) \}
 \nonumber \\
\Gamma^{gf}_{\alpha}(\omega)&=& [\Gamma^{fg}_{\alpha}(\omega)]^*=
\mbox{sg}(\omega) \frac{|w_{c, \alpha}|^2}{2 w_{\alpha}^2} \frac{\Delta_{\alpha}}{r(\omega)} \nonumber \\
& &
\times [s^{+}(\omega)+s^{-}(\omega)],
\end{eqnarray}
with $r(\omega)=\Theta(|\omega|-|\Delta_{\alpha}|)\sqrt{\omega^2-\Delta_{\alpha}^2}$ and $s^{\pm}(\omega)=
\Theta(|2w_{\alpha}|-|r(\omega) \pm \mu|)
\sqrt{4w_{\alpha}^2-(r(\omega) \pm \mu)^2}$.
It can be verified that, for $\mu=0$, $\Gamma^{gg}_{\alpha}(\omega)$
reduces to the $\mbox{Im}$ of the diagonal component of the self-energy 
defined by an infinite tight-binding wire with local pairing 
reported in Ref. \onlinecite{vec}.

The final expressions for the retarded self-energies in the thermodynamic limit can be obtained
by recourse to the Kramers-Kronig relation:
\begin{equation}
\Sigma^{\nu \nu', R}(\omega)=\int_{-\infty}^{+\infty} \frac{d \omega'}{2 \pi} \frac{ \Gamma^{\nu,\nu'}(\omega')}{\omega - \omega'+ i \eta}
\end{equation}.

\section{Dyson's equation for $\hat{G}^<_{\sigma}$ and  $\hat{F}^<_{\sigma}$.}
The lesser counterpart of (\ref{retc}) is:
 \begin{eqnarray} \label{less}
& & [\hat{1} \omega - \hat{\Sigma}^{gg,R} (\omega) - \hat{\varepsilon} ]
\hat{G}^<_{\sigma}(\omega)
+ \hat{\Sigma}^{gf,R} (\omega) \hat{F}^<_{\sigma}(\omega)   \nonumber \\
& & =
\hat{\Sigma}^{gg,<} (\omega) \hat{G}^A_{\sigma}(\omega)- 
\hat{\Sigma}^{gf,<} (\omega) \hat{F}^A_{\sigma}(\omega), \nonumber \\
& &  [\hat{1} \omega - \hat{\Sigma}^{ff,R} (\omega) + \hat{\varepsilon} ]
\hat{F}^<_{\sigma}(\omega)
+ \hat{\Sigma}^{fg,R} (\omega) \hat{G}^<_{\sigma}(\omega)  \nonumber \\
& &  = 
\hat{\Sigma}^{ff,<} (\omega) \hat{F}^A_{\sigma}(\omega)
- \hat{\Sigma}^{fg,<} (\omega) \hat{G}^A_{\sigma}(\omega).
\end{eqnarray}

\end{document}